\documentclass[preprint2]{aastex}

\slugcomment{Submitted to Astronomical Journal in September 2012}

\shorttitle{Variability of T Tauri Stars}
\shortauthors{Chou et
al.}

\begin{document}

\bibliographystyle{astron}

\title{Time Variability of Emission Lines for Four Active T Tauri Stars (I): October--December in 2010\footnote{Based on
observations obtained at the Canada-France-Hawaii Telescope (CFHT)
which is operated by the National Research Council of Canada, the
Institut National des Sciences de l'Univers of the Centre National
de la Recherche Scientifique of France, and the University of
Hawaii.}}

\author{Mei-Yin Chou\altaffilmark{1}, Michihiro Takami\altaffilmark{1}, Nadine Manset\altaffilmark{2},
Tracy Beck\altaffilmark{3}, Tae-Soo Pyo\altaffilmark{4}, Wen-Ping
Chen\altaffilmark{5}, Neelam Panwar\altaffilmark{5}, Jennifer L.
Karr\altaffilmark{1}, Hsien Shang\altaffilmark{1}, Hauyu Baobab
Liu\altaffilmark{1} }

\altaffiltext{1}{Institute of Astronomy and Astrophysics, Academia
Sinica, P.O. Box 23-141, Taipei 10617, Taiwan}
\altaffiltext{2}{Canada-France-Hawaii Telescope, 65-1238 Mamalahoa
Hwy, Kamuela, HI 96743, USA} \altaffiltext{3}{The Space Telescope
Science Institute, 3700 San Martin Dr. Baltimore, MD 21218, USA}
\altaffiltext{4}{Subaru Telescope, 650 North Aohoku Place, Hilo,
HI 96720, USA} \altaffiltext{5}{Institute of Astronomy, National
Central University, Taiwan 320, Taiwan}

\begin{abstract}
We present optical spectrophotometric monitoring of four active T
Tauri stars (DG Tau, RY Tau, XZ Tau, RW Aur A) at high spectral
resolution ($R \ga 1 \times 10^4$), to investigate the correlation
between time variable mass ejection seen in the jet/wind structure
of the driving source and time variable mass accretion probed by
optical emission lines. This may allow us to constrain the
understanding of the jet/wind launching mechanism, the location of
the launching region, and the physical link with magnetospheric
mass accretion. In 2010, observations were made at six different
epochs to investigate how daily and monthly variability might
affect such a study. We perform comparisons between the line
profiles we observed and those in the literature over a period of
decades and confirm the presence of time variability separate from
the daily and monthly variability during our observations. This is
so far consistent with the idea that these line profiles have a
long term variability (3-20 years) related to episodic mass
ejection suggested by the structures in the extended flow
components. We also investigate the correlations between
equivalent widths and between luminosities for different lines. We
find that these correlations are consistent with the present
paradigm of steady magnetospheric mass accretion and emission line
regions that are close to the star.

\end{abstract}

\keywords{stars: pre-main-sequence ---  stars: activity --- stars:
emission-line, Be --- stars: winds, outflows --- stars: individual
(DG Tau, RY Tau, XZ Tau, RW Aur A) --- line: profiles
--- accretion, accretion disks}

\section{Introduction}

Classical T Tauri stars (CTTSs) are low-mass pre-main sequence
stars at the evolutionary stage when the surrounding accretion
disks become optically visible, and firstly discovered by
\citet{Joy45}. CTTSs are distinguished by their rich permitted
line emission which is produced mainly in the magnetic funnel flow
originated from the inner disk edge \citep[e.g.,][]{Muzerolle01},
and an additional optical/UV excess emission continuum (veiling)
produced in ''hot spots'' where the accreted material impacts the
stellar surface \citep[][]{Calvet98}. The permitted lines often
show inverse P Cygni profiles which may indicate funnel flows by
the magnetospheric mass accretion \citep[][for review]{Najita00}.

Moreover, many of the CTTSs host collimated jets
\citep[e.g.,][]{Hartigan95, Eisloffel00}. Understanding their
driving mechanism and their physical link with protostellar
evolution are the most important issues in star formation
theories. There is growing evidence that these jets are powered by
disk accretion \citep{Cabrit90,Hartigan95,Calvet97}.
High-resolution observations have revealed internal flow motion
consistent with rotation. This agrees with the idea that the jet
removes excess angular momentum from the accreting material
\citep{Bacciotti02,Coffey04,Coffey07}, but more detailed
investigations are required for confirmation \citep[see][for
ongoing projects]{Cabrit06,Coffey12}.

There are several theories to explain the driving mechanism of the
jet and its physical link with mass accretion. The
magneto-centrifugal wind models, including X-wind \citep{Shu00}
and disk wind \citep{Konigl00} models, are regarded as the most
promising. The former proposes that the jet launches from the
inner edge of the disk (within 0.1 AU of the star), while the
latter proposes that the jet launching region covers a wider disk
surface at a few AU scales. Furthermore, magnetic pressure and
reconnection wind models are also candidates
\citep[e.g.,][]{Goodson99}, although the reconnection wind could
be the non-steady state of breaking a magnetosphere which is
similar to the X-wind. In this context, studies of jets are
important for understanding not only star formation, but also
circumstellar disks and the ongoing (or initial conditions of)
planet formation. However, the limited angular resolutions of
present telescopes are far from sufficient for resolving the jet
launching or flow acceleration regions, which are predicted to be
located within a few AU from the star.

Alternatively, simultaneous monitoring of jet structures and
signatures of magnetospheric accretion would allow us to test
theories of jet driving and mass accretion. Over the last decade,
observations at high-angular resolutions have allowed us to reveal
fine structures of jets and winds close to the star at $\ga10$ AU,
measure their proper motions and investigate their episodic mass
ejection with a time scale of 3--20 years
\citep[e.g.,][]{Pyo03,Lopez03,Krist08,Agra11}. Permitted emission
lines, i.e., probable signature of mass accretion, are also known
to show time variability
\citep[e.g.,][]{Petrov90,Petrov99,Petrov01a,Johns95,Alencar05,Mendigutia11a}.
If the jet is driven by an X-wind as well as a reconnected wind,
the emergence of new knots in the extended jets should be well
correlated with the time variability of magnetospheric accretion
because both the jet launching and material accretion occur on the
inner edge of the disk. If the jet is driven by a disk-wind, such
a correlation may be weak since the disk wind and the
magnetospheric accretion are not directly related to each other.
As such, we should be able to test theories of jets, wind and mass
accretion.

We have thus started a monitoring project for four well studied
active T Tauri stars: DG Tau, RY Tau, XZ Tau and RW Aur A (see
Appendix for details of individual targets). These exhibit a
variety of optical emission/absorption lines (see Section 3.2 for
previous studies), and the extended jet or wind at a $\sim0$''.1
resolution (see Appendix for a review). In this paper we present
optical high-resolution spectroscopy for the first year
observations of 2010. It is known that active CTTSs show
short-term (daily, monthly and yearly) variabilities in line
fluxes, line profiles and veiling continuum
\citep[][]{Bouvier93,Bouvier95,Mendigutia11a}. The daily
variability for many CTTSs is due to a combination of stellar
rotation \citep[typically $P=2$ to 20 days,][]{Bouvier93,Choi96}
and the non-uniform distribution of magnetospheric channel flow
and the resultant hot spots on the stellar surface
\citep[][]{Bouvier93,Bouvier95}. Although cool spots similar to
sunspots may also contribute to the variability, the effects are
more prominent in weak-line T Tauri stars than in CTTSs
\citep[e.g.,][]{Herbst94}. In addition, the patterns of cool spots
usually last for several periods or longer so their contribution
to the daily variability may be unapparent. In contrast, active
CTTSs may show such a short-term variability due to time-variable
mass accretion as well as a long-term variability. An
understanding of this variability for the target stars is
necessary if we want to be able to detect variability related to
episodic mass ejection over the other variabilities.

In Section 2 we describe the observations and data reduction. In
Section 3 we show the line profiles, and investigate their time
variabilities during our observations in comparison of those in
the literature over decades. In Section 4 we show the equivalent
widths and luminosities, the correlations, and comparisons with
the continuum flux. Discussion and summary are described in
Sections 5 and 6, respectively. Throughout the paper we assume a
distance of 140 pc of our targets \citep{Wichmann98}.

\section{Observations and Data Reduction}

We obtained high resolution spectra using the 3.6-m
Canada-France-Hawaii Telescope (CFHT) with ESPaDOnS, CFHT's
Echelle spectropolarimetric device for the observation of stars,
on six nights: October 21, November 17, 21, 25, 27 and December 17
in 2010. The observations were made using the ''object+sky
spectroscopy mode'', for which the spectra of the object and sky
are simultaneously obtained through different fibers. The spectra
cover the wavelength range $\sim3,700-10,500$ \AA\ with a
resolution of $68,000$. This large spectral coverage of ESPaDOnS
includes a number of emission lines. In this paper we present our
study on the following permitted lines studied by
\citet{Muzerolle98_opt} for 11 T Tauri stars: H$\alpha$ 6563 \AA,
Pa 11 8863 \AA, \ion{Ca}{+2} 8498/8542/8662 \AA\ triplet,
\ion{He}{+1} 5876/6678/7065 \AA, \ion{O}{+1} 7772/8446 \AA, and Na
D 5890/5896 \AA. We also show three forbidden lines which are
formed in the jet regions for comparison: [\ion{O}{+1}] 6300 \AA,
[\ion{S}{+2}] 6731 \AA\ and [\ion{Fe}{+2}] 7155 \AA.

All four targets and two standard stars were observed each night
consecutively and within a $\sim30$ minutes interval. These were
made with photometric conditions with the exception of December
17. The line and continuum luminosities are derived from the data,
except for the December 17 observations, which were not obtained
under stable sky transparency. The average signal-to-noise ratios
at 8090 \AA\ are $\sim44$, $\sim61$, $\sim67$, and $\sim37$ for DG
Tau, RY Tau, XZ Tau and RW Aur, respectively. Data were processed
using the standard data reduction pipeline Upena provided by the
CFHT, which is based on the Libre-ESpRIT package \citep{Donati97}
, and the sky subtraction has been performed in our spectra.
Furthermore, we have remove telluric absorption for the
[\ion{O}{+1}] 6300 \AA\ line in the spectra. The systemic velocity
for each target star on each observation date was calibrated using
photospheric absorption lines \ion{Li}{+1} 6708 \AA\ and
\ion{Ca}{+1} 6439 \AA\ for DG Tau, RY Tau, and XZ Tau. The stellar
photospheric absorption of RW Aur A was too shallow for accurate
measurements of the velocity so we instead used the velocity by
\citet{Folha01}, who tabulate the systemic velocity for a number
of T Tauri stars including our four targets.

The large aperture of ESPaDOnS (1''.6 in diameter) compared with
the seeing (0''.4--1''.0 during our observations) allows
spectrophotometry. Two standard stars, HD 283642 and HD 42784,
were also observed for each date for flux calibration. For
accurate calibration at each line luminosity we have also made
$VRI$ photometry of these two standard stars using the Tenagra
80-cm telescope in Arizona, USA, on November 17 and December 08,
2011. The measured magnitudes are listed in Table \ref{std_BVRI}.
The line fluxes of targets were calibrated by interpolating the
photometric data to the target wavelength and comparing the
derived fluxes with the photon counts in the ESPaDOnS spectra of
the standard stars. The derived fluxes with two standard stars
have discrepancies in the flux conversion factor up to $\sim$20,
$\sim$30, $\sim40\%$ at $VRI$ wavelengths (5500, 7100, and 9700
\AA, respectively). We thus take the average of the two conversion
factors obtained from two different standard stars, resulting the
uncertainty of the absolute flux calibration to be up to $\sim$10,
$\sim$15 and $\sim20\%$ at 5500, 7100, and 9700 \AA, respectively.

RW Aur A has a companion with a separation of 1''.4
\citep[e.g.,][]{Ghez93,Leinert93}. Our spectra obtained on October
21 and December 17 are suspected to have a contribution from the
companion due to the relatively poor seeing on these dates
($\sim$1''.0). Therefore, these spectra are not used in this
paper. XZ Tau is also associated with a companion, and due to its
small separation (0''.3) the spectra of both binary components
were included in the ESPaDOnS aperture. However, the XZ Tau S
component has often been brighter to much brighter than the 0''.3
separation companion at optical wavelengths and the emission line
profiles are assumed to be applicable mostly to XZ Tau S. See
Appendix for the details of confirmed/possible companions for
individual objects.

\section{Line Profiles}

Figures 1--4 show the observed line profiles at the six epochs.
The spectra have been adjusted to the continuum level and smoothed
using a moving average over 15 data points of the reduced spectra
for better signal-to-noise; the resolution is $\sim30$ km s$^{-1}$
after smoothing. We describe the time variation of the permitted
lines during our observations in Section 3.1, and differences from
those in previous publications in Section 3.2.

\subsection{Variation of Permitted Line Profiles in 2010}

\subsubsection{DG Tau, RW Aur A}

The observed line shapes do not vary much with the exception of
the line-to-continuum ratio. In DG Tau, there is a weak redshifted
absorption in the \ion{Ca}{+2} triplet (at $v\sim50$ km s$^{-1}$)
on November 21 which is not apparent in other nights.

In RW Aur A, the \ion{O}{+1} 8446 \AA\ line profile is flatter at
the peak on November 17 and 21 than the other two dates, and
appears to be associated with shallow absorption at zero velocity.
In the Na D profiles, the velocity range of the absorption is
$\sim$--150 to $\sim$0 km s$^{-1}$ in November 25 and 27, while
the profiles of November 17 and 21 are also associated with
shallow absorption extending toward the positive velocity up to
$\sim$150 km s$^{-1}$.

\subsubsection{RY Tau}

In H$\alpha$, the redshifted emission components show a larger
variation in the line-to-continuum ratio (by a factor of 2) than
the blueshifted emission ($\sim30\%$). The profile of November 27
shows absorption at $v\sim100$ km s$^{-1}$.

The \ion{Ca}{+2} emission shows slight redshifted absorption on
October 21 and November 25, with slightly blueshifted absorption
on November 17 (at $v\sim-50$ km s$^{-1}$), and absorption near
the zero velocity on November 21 and 27.

For \ion{O}{+1} 7772 \AA, the line shapes on October 21, November
17 and 21 show two absorption features (near the zero velocity and
at $v\sim100$ km s$^{-1}$) with small emission features on both
sides. In contrast, the line shapes on November 25 and 27 show
strong absorption ($\sim200$ km s$^{-1}$ in FWHM) and shallow
emission features on the blueshifted side only.

\subsubsection{XZ Tau}

The H$\alpha$ profile on October 21 shows a blueshifted absorption
at $\sim-100$ km s$^{-1}$. On October 21 and November 17 the
H$\alpha$ and \ion{Ca}{+2} lines show marginal redshifted
absorption at $\sim50$ km s$^{-1}$, and the \ion{He}{+1} lines
show stronger central emission. The Pa 11 and \ion{He}{+1} 7065
\AA\ emission features were absent on December 17. On the same
date the blueshifted absorption in the Pa 11 line shows a larger
wing-like absorption feature toward the zero velocity.

\subsection{Differences from Previous Observations}

The previous observations of our target lines are summarized in
Table \ref{literature}. While some line profiles are similar to
ours, we find a number of differences which are described below
for DG Tau, RY Tau, and RW Aur A. We limit our comparison to the
shape of the permitted lines only.

\subsubsection{DG Tau}

Our H$\alpha$ line profiles show a peak at $\sim-50$ km s$^{-1}$,
a dip (or an emission minimum) at $\sim-30$ km s$^{-1}$ and
marginal absorption at $\sim$70 km s$^{-1}$. These are not seen in
the profile observed in 1981 by \citet{Mundt84}, at three epochs
in 1988--1992 by \citet{Beristain01}, or in 1996 by
\citet{Muzerolle98_opt}. A blueshifted absorption line at
$v\sim-50$ km s$^{-1}$ is observed only in our spectra, and two
epochs shown in \citet{Beristain01}.

\begin{figure*}
\vspace{-3cm} \epsscale{2.1} \plotone{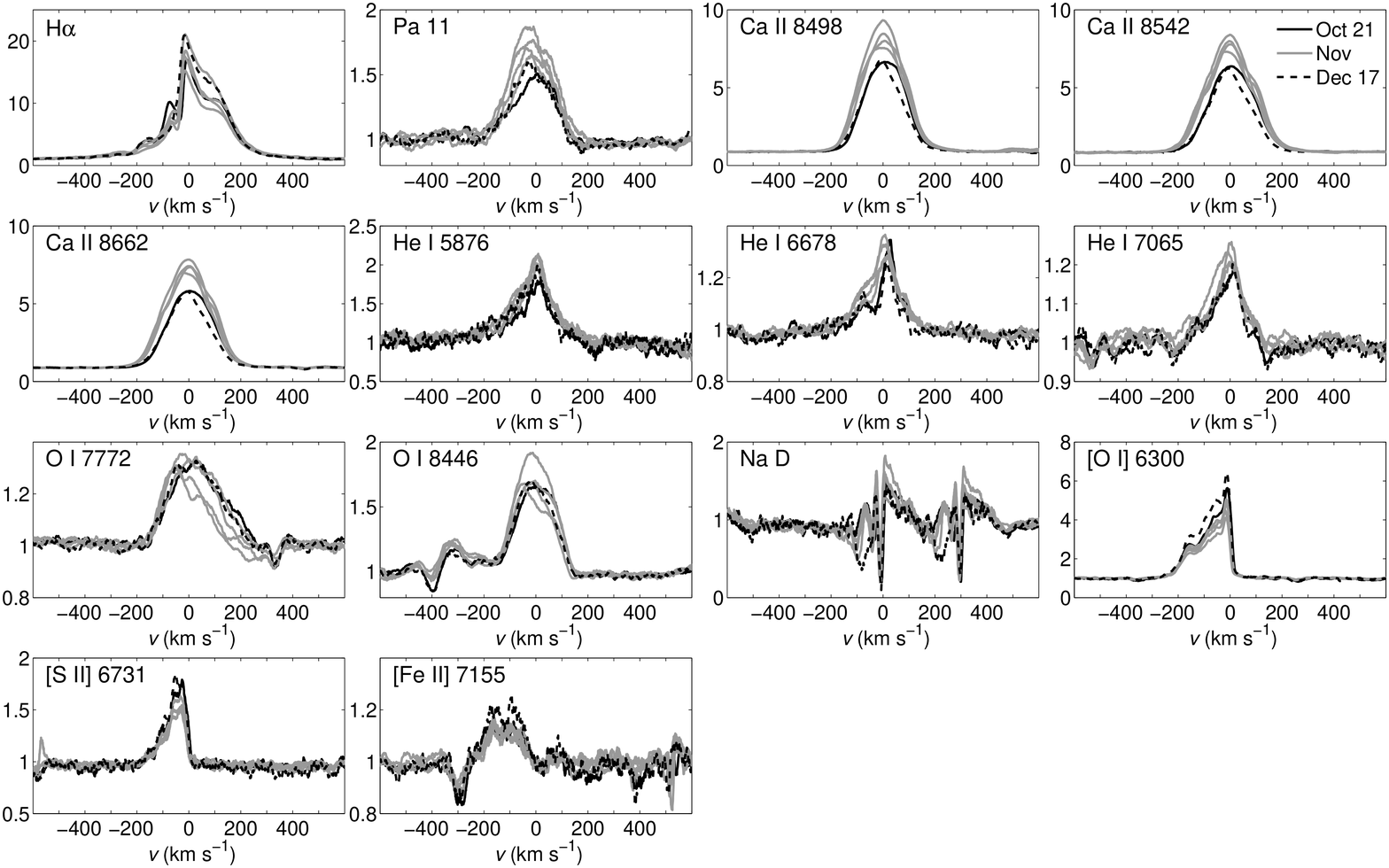} \caption{Line
profiles observed in DG Tau. For each line the plot shows the
profiles observed on six nights in October to December 2010. For
all the profiles the flux is normalized to the continuum, and the
velocity is shown in term of stellar velocity.
\label{DGTau_profiles}}
\end{figure*}

The \ion{O}{+1} 7772 \AA\ line observed in 1996
\citep{Muzerolle98_opt} shows four peaks at $\sim-150$, 0, 150 and
350 km s$^{-1}$. In contrast, our spectra show a single peak
without any absorption on most of the nights observed, except for
November 21, where it shows weak absorption near the zero
velocity. The \ion{O}{+1} 8446 \AA\ line observed in 1996
\citep{Muzerolle98_opt} showed a double-peak feature which may has
been caused by shallow absorption at zero velocity. This
absorption is seen only marginally (if not absent) in the profile
observed in 1987 \citep{Hamann92a} and in Figure
\ref{DGTau_profiles}.

The Na D profiles observed in 1981 \citep{Mundt84} are associated
with broader and shallower blueshifted absorption at $v=-200$ to
$-100$ km s$^{-1}$.

The H$\alpha$ and \ion{He}{+1} 5876 \AA\ profiles in
\citet{Mundt84,Beristain01} shows another emission component at
$v\sim-250$ km s$^{-1}$, corresponding to the extended jet (see
Appendix for references). This may not be due to the intrinsic
variability of the emission line regions, since the observed
brightness should also depends on the slit width and position (or
the size and position of the aperture), and also on the seeing.

\subsubsection{RY Tau}

\begin{figure*}
\epsscale{2.1} \plotone{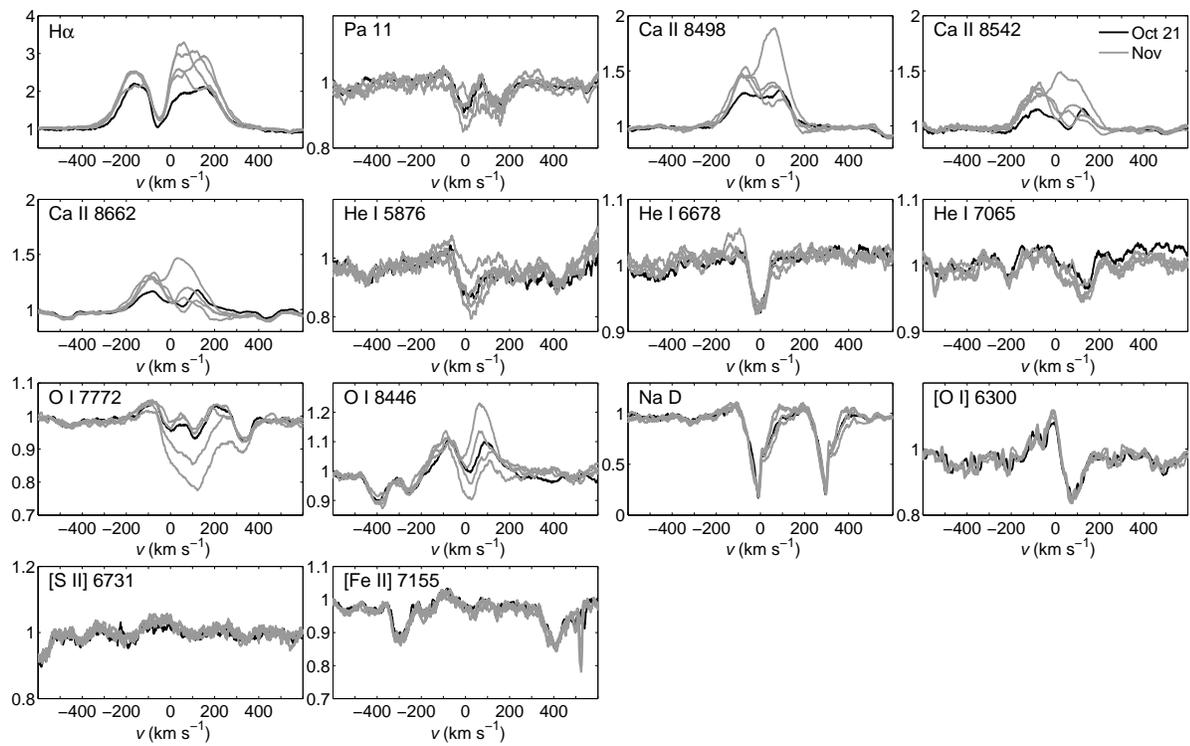} \caption{Same as Fig.
\ref{DGTau_profiles} but for RY Tau. \label{RYTau_profiles} }
\end{figure*}

The H$\alpha$ profiles in Figure \ref{RYTau_profiles} and those in
most of the literature show two distinct emission components at
blueshifted and redshifted velocities, with deep absorption at
zero velocity. The brightness of these two components are
comparable in Figure \ref{RYTau_profiles} and most of the profiles
in the literature. However, the blueshifted emission is
significantly fainter than the redshifted emission in several
epochs in 1980--1998
\citep{Mundt82,Zajtseva85,Petrov90,Mendigutia11a}. The blueshifted
emission is marginal, if not absent, in one of the profiles
observed in 1984 \citep{Zajtseva85}; this profile instead shows
shallow blueshifted absorption at $-1000$ to $-500$ km s$^{-1}$.
\citet{Zajtseva85} and \citet{Petrov90} show that such a variation
occurred within 1--2 months.

Most of the H$\alpha$ profiles show a deep absorption feature at
the zero velocity in which the line-to-continuum ratio $\sim0$.
However, the emission level at this level is close to the peak of
either the blueshifted or redshifted emission in 1984
\citep{Edwards87}. Furthermore, of all the profiles, the shallow
absorption at the peak of the redshifted emission is observed only
in our spectra, in 1984 \citep{Edwards87}, and possibly two epochs
in 1988 \citep{Petrov90}.

The \ion{Ca}{+2} profiles observed in 1987 \citep{Hamann92a} are
similar to our observed features from November 17, in which the
redshifted emission shows a brighter peak than the blueshifted
one. In contrast, those observed in 1989--1996 \citep{Petrov99}
are significantly different from any of the other profiles. The
former consist of  two emission peaks at $v \sim -150$ and 150 km
s$^{-1}$ and absorption at the zero velocity with the bottom of
the absorption 20--30\% below the continuum level during their
observations.

Some Na D profiles observed in 1987--1996
\citep{Petrov90,Petrov99} appear to be associated with marginal
additional blueshifted and/or redshifted absorption at the
threshold of the absorption feature, making the entire absorption
broader. The Na D profiles observed in January 1999
\citep{Mendigutia11a} show a remarkable emission component on the
blueshifted side.

\subsubsection{RW Aur A}

The peak of H$\alpha$ profiles in the redshifted emission is
brighter than that of the blueshifted emission in most of the
profiles, but nearly the same for a single epoch in 1988--1992
\citep{Beristain01} and on November 25, 2010 (Fig.
\ref{RWAur_profiles}). The redshifted emission is slightly fainter
than the blueshifted emission in 1981 \citep{Mundt84}. Most of the
line profiles, including those in Figure \ref{RWAur_profiles},
show a velocity $\sim-50$ km s$^{-1}$ at the bottom of the central
absorption. In contrast, it is $\sim-100$ km s$^{-1}$ in some
profiles in 1989--1999
\citep{Muzerolle98_opt,Petrov01a,Alencar05}. Some of the Pa 11
profiles in 1998 \citep{Petrov01a} show a blueward asymmetry. This
contrasts to the other profiles in the literature and Fig.
\ref{RWAur_profiles}, which show a symmetric profile about the
stellar velocity.

The \ion{Ca}{+2} profiles in Figure \ref{RWAur_profiles}, in
particular those from November 21 and 25, are similar to most of
those in 1986--1999 \citep{Hamann92a,Petrov01a,Alencar05}. Our
profiles from November 17 and 27 are similar to that of 1996
\citep{Muzerolle98_opt}, but the peak of the redshifted emission
shows an asymmetry in the latter, perhaps due to the presence of
shallow absorption. Two profiles observed in 1986--1987
\citep{Alencar05} are remarkably different from the others. Both
are nearly symmetric about the stellar velocity, and one of them
is triangular while the other is flat-topped.

\citet{Beristain01} show that the He lines in CTTSs often have a
broad component (BC, FWHM$=130-300$ km s$^{-1}$) and a narrow
component (NC, FWHM$=30-60$ km s$^{-1}$) which are approximately
symmetric about the zero velocity. All of the \ion{He}{+1} 5876
\AA\ lines in the literature in Table \ref{literature} clearly
show both the BC and NC. In contrast, our profiles do not clearly
show the presence of a NC.

Some of the \ion{O}{+1} 7772 \AA\ profiles observed in 1995--1999
\citep{Petrov01a} are significantly different from the others.
These are dominated by an emission feature whose peak is
blueshifted at $\sim-150$ km s$^{-1}$. Some of them show a
redshifted absorption without blueshifted emission. The
\ion{O}{+1} 8446 \AA\ line in 1996 \citep{Muzerolle98_opt} shows a
redshifted absorption at $200-350$ km s$^{-1}$.

Some Na D profiles observed in 1989--1999
\citep{Petrov01a,Alencar05} are associated with remarkable
emission over the continuum level at the both sides of the
absorption. Those in 1980 \citep{Mundt84} and 1996
\citep{Muzerolle98_opt} show an inverse P-Cygni profile.

\section{Equivalent Widths and Luminosities}

We measure the equivalent widths and luminosities of fourteen
lines for each target. These are tabulated in Tables \ref{EWs} and
\ref{line_luminosities}, respectively. The latter also includes
the continuum luminosities at 5000 and 8000 \AA, and has been
extinction-corrected. The following discussion is only for
permitted lines since they are formed in the near-stellar region
which is the focus of this study. For each target, changes in
equivalent widths are remarkably different between lines. In DG
Tau, the change is 45--50\% for most of the emission lines but
only 12 and 21\% in \ion{He}{+1} 5876 and 6678 \AA, respectively;
in RY Tau, it is a factor of $\sim$4 for \ion{O}{+1} 8446 \AA\ but
$\sim$1.5 in H$\alpha$; in XZ Tau, it is a factor of $\sim$4 for
the \ion{Ca}{+2} triplet but only 7\% for H$\alpha$; in RW Aur A,
the change is $\sim40\%$ in \ion{Ca}{+2} triplets but $\sim20\%$
for H$\alpha$. Among the emission features observed, \ion{O}{+1}
8446 \AA\ and the \ion{Ca}{+2} triplet tend to show a large
variation in equivalent width. In contrast, \ion{He}{+1} and
H$\alpha$ are the most stable lines. The equivalent widths for RY
Tau are usually smaller than other stars except for the
\ion{O}{+1} 7772 \AA\ line, in which RW Aur A has stronger
absorption feature than RY Tau.

Table \ref{literature} shows the equivalent widths of H$\alpha$
observations from the literature. In DG Tau, the equivalent widths
during our observations (74--107 \AA) are significantly larger
than those measured by \citet{Muzerolle98_opt} in 1988--1992 (46
\AA) and smaller than those measured by \citet{Mundt84} (110 \AA)
in 1981. For RY Tau and RW Aur A, i.e, targets for which optical
spectroscopy has been extensively made, the equivalent widths we
observed (10--15 and 65--80 \AA, respectively) are within the
ranges in the literature (6--25 and 50--123 \AA, respectively). It
is intriguing that the equivalent widths measured by
\citet{Fernandez95} varied by a factor of 2 in the four days of
their observations (September 15--19, 1989), while those in our
data show only $\sim20\%$ variability.

\begin{figure*}
\epsscale{2.1} \plotone{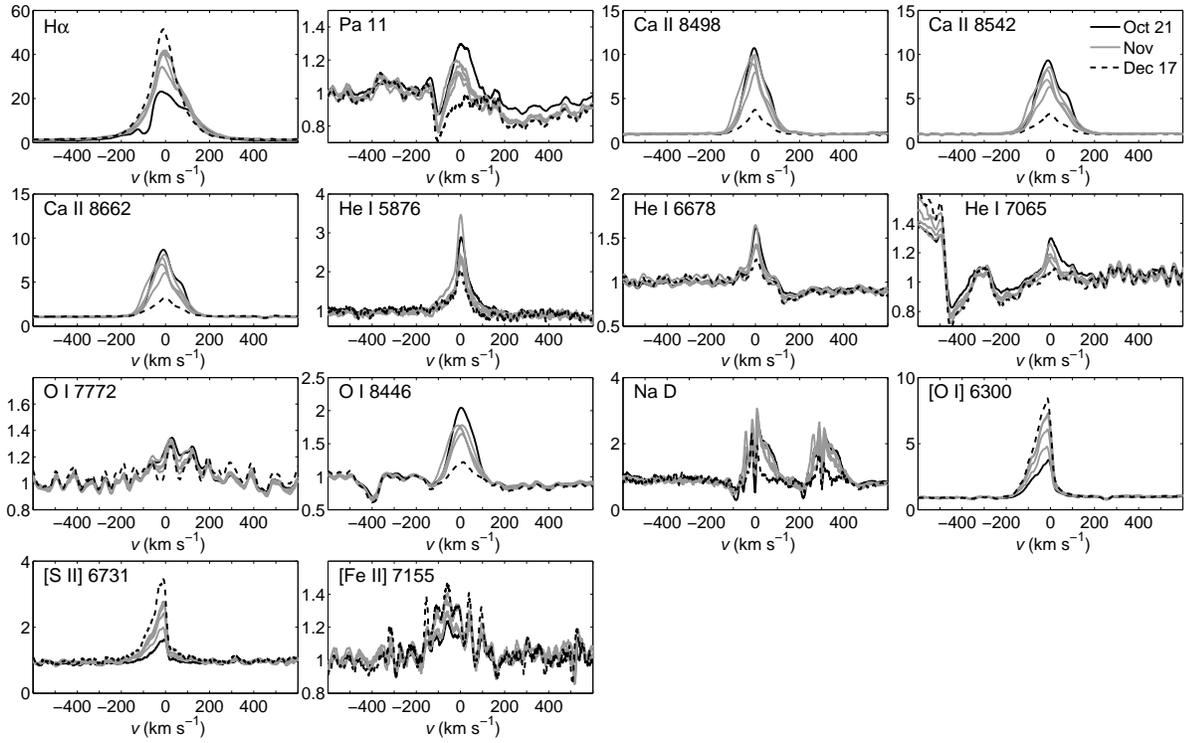} \caption{Same as Figs
\ref{DGTau_profiles}--\ref{RYTau_profiles} but for XZ Tau. Ripple
patterns in \ion{O}{+1} 7772 \AA\ line may be due to the TiO band.
\label{XZTau_profiles}}
\end{figure*}

\begin{figure*}
\epsscale{2.1} \plotone{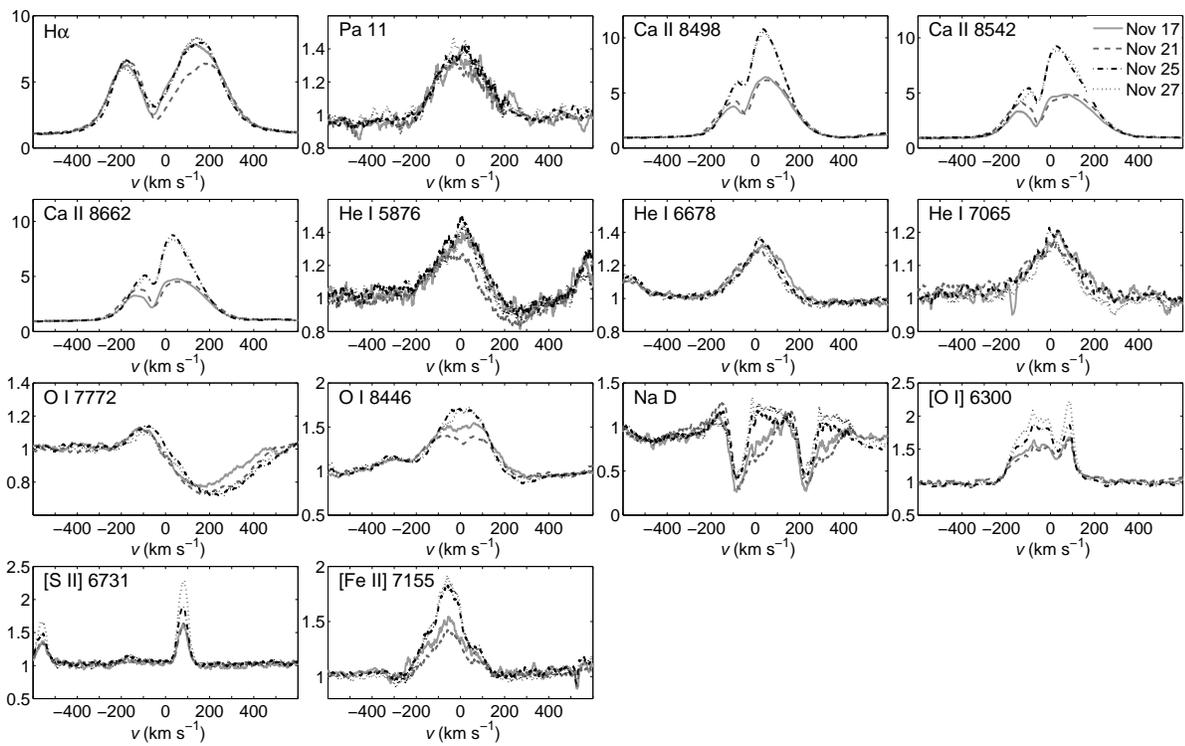} \caption{Same as Figs
\ref{DGTau_profiles}--\ref{XZTau_profiles} but for RW Aur A.
\label{RWAur_profiles}}
\end{figure*}

The luminosity varies in H$\alpha$ and in those lines without
absorption by a factor of $\sim 2$ for each star. A larger
variation is observed for XZ Tau on Dec 17, on which the line
luminosities are lower than the other dates by a factor of up to
4--13 for the \ion{He}{+1} 5876\AA\ and \ion{O}{+1} 8446 \AA\ line
and the \ion{Ca}{+2} triplet. Similar but more marginal decrease
is also seen in line-to-continuum ratios in Figure
\ref{XZTau_profiles} and equivalent widths in Table \ref{EWs}.

\begin{figure*}
\epsscale{2.1} \plotone{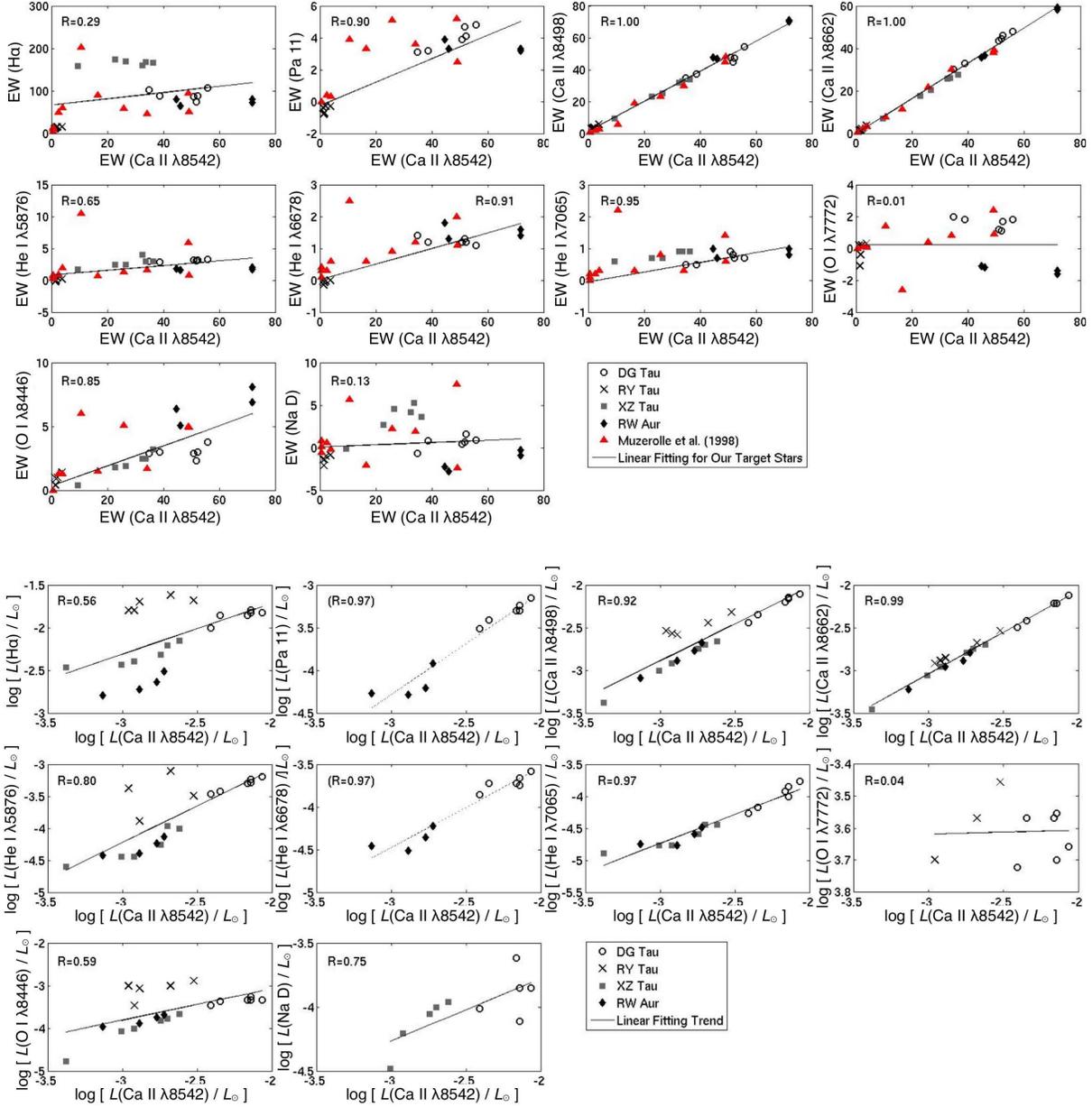} \caption{(upper) Correlations for
equivalent widths between \ion{Ca}{2} 8542 \AA\ and the other
permitted lines. Positive values denote emission, and negative
values denote absorption. The $R$ value in each panel represents
the correlation coefficient, and the solid line represents the
linear regression for our target stars. The triangle symbols are
the measurements for 11 CTTSs by \citet{Muzerolle98_opt} for
comparison. (lower) Same as the upper plots but for the
extinction-corrected line luminosity. Only those with positive
equivalent widths are used. In the case of the correlations
between Pa 11, \ion{He}{+1} 6678 \AA\ and \ion{Ca}{+2} 8542 \AA,
the linear regression and correlation coefficients are tentative
as this is due to the difference between only two objects being
plotted.} \label{corr_line_vs_line}
\end{figure*}

Figure \ref{corr_line_vs_line} shows the correlation of equivalent
widths and extinction-corrected line luminosities between
\ion{Ca}{+2} 8542 \AA\ and the other 10 lines. We select
\ion{Ca}{+2} 8542 \AA\ for reference since this line is known to
show a good correlation with accretion luminosity
\citep{Muzerolle98_opt,Calvet00}. Among the 10 lines, the other
two \ion{Ca}{+2} lines show an excellent correlation with
\ion{Ca}{+2} 8542 \AA. Three \ion{He}{+1} lines, \ion{O}{+1} 8446
\AA, and Pa 11 also have a fairly strong correlations with
\ion{Ca}{+2} 8542 \AA. There is little to no correlation for Na D,
\ion{O}{+1} 7772 \AA, and H$\alpha$ lines, which have apparent
absorption features in most of the target stars. The upper panel
of Figure \ref{corr_line_vs_line} also shows the equivalent widths
measurements for 11 T Tauri stars by \citet{Muzerolle98_opt}. The
measurements between \citet{Muzerolle98_opt} and our observation
show similar correlations between \ion{Ca}{+2} 8542 \AA\ and the
other lines, except for a single object (BP Tau) in the
\ion{He}{+1} lines.

\begin{figure*}
\epsscale{2.1} \plotone{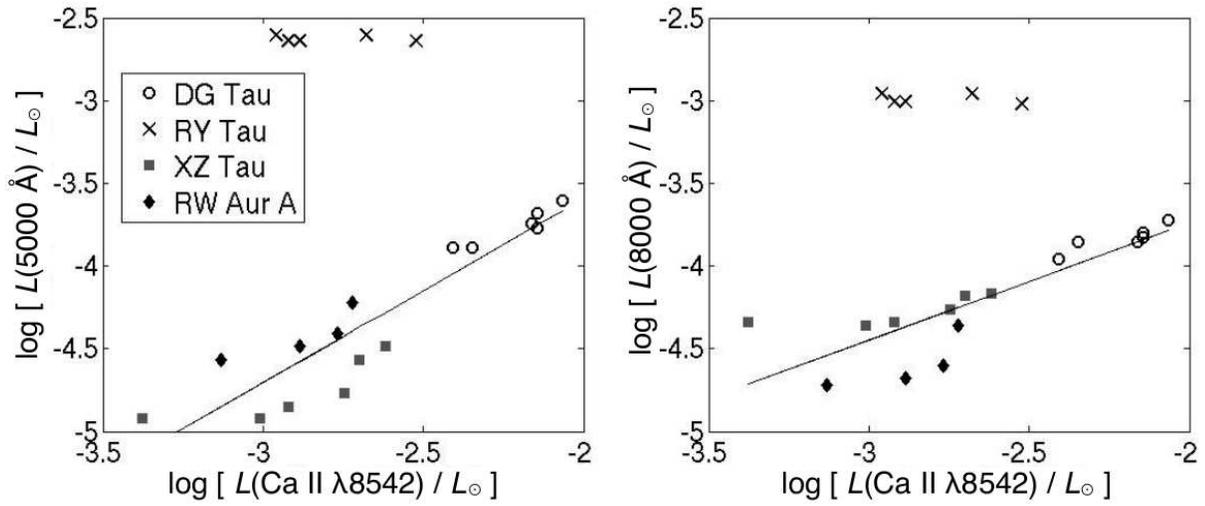} \caption{ The correlation between
the extinction-corrected line luminosity for \ion{Ca}{+2} 8542
\AA\ and continuum luminosity at 5000 (left) and 8000 (right) \AA.
The regression line is determined for three objects (DG Tau, XZ
Tau and RW Aur A) in both figures. } \label{corr_line_vs_cont}
\end{figure*}

Figure \ref{corr_line_vs_cont} shows the extinction-corrected
luminosity of the continuum level compared with the \ion{Ca}{+2}
8542 \AA\ line for the four target stars. In DG Tau, XZ Tau and RW
Aur A the continuum luminosity increases with the \ion{Ca}{+2}
8542 \AA\ luminosity. These are fit well with a single regression
line at 5000 and 8000 \AA, respectively. The 5000 \AA\ continuum
shows a larger increase (by a factor of $\sim1.3$ dex over
$\sim1.3$ dex for the \ion{Ca}{+2} 8542 \AA\ luminosity) than 8000
\AA\ ($\sim1.0$ dex). In contrast to these three objects the
continuum luminosity remained constant in RY Tau both at 5000 and
8000 \AA\ despite the presence of variation in the \ion{Ca}{+2}
8542 \AA\ luminosity.

\section{Discussion}

\subsection{Consistency with The Present Paradigm for Magnetospheric Mass Accretion}

It is believed that, in many cases, permitted line emission
associated with CTTSs primarily originates from magnetospheric
accretion columns \citep[e.g.,][for a review]{Najita00}. Although
their complicated geometry has been extensively discussed over the
past decade \citep[see][for a review]{Bouvier07_PPV}, models with
a simplified geometry (i.e., with stellar dipole magnetic field
axisymmetric about the stellar rotation axis) have been successful
in explaining many line profiles and their luminosities
\citep[e.g.,][]{Muzerolle01}. The accretion flow originates from
the inner disk edge, and hits the stellar surface at a few hundred
km s$^{-1}$. This heats up the stellar surface and creates ''hot
spots'' observed as an excess in the UV continuum. The
distribution of such hot spots is not usually uniform, and as a
result, this causes periodic time variability corresponding with
the stellar rotation. This has been extensively observed over
decades to measure the rotational period of many T Tauri stars
\citep[see][for a review]{Herbst07}. This scenario also explains
the good correlation between the accretion luminosity measured
using the UV excess and luminosities of permitted line such as the
\ion{Ca}{2 } triplet \citep{Muzerolle98_opt,Calvet00}. The
collision of the accretion flow on the stellar surface also causes
accretion shock toward the accretion flow. This explains the
presence of NC in addition to BC in some emission line profiles
such as \ion{He}{+1} (Section 3.1).

In contrast, the contribution from the inner wind may be
significant for Balmer lines in active CTTSs
\citep[e.g.,][]{Takami01,Takami03,Alencar05,Kurosawa06}. The Na D
lines in many objects and Balmer lines in some objects show
blueshifted absorption. This indicates the presence of such a wind
with velocity up to a few hundred km s$^{-1}$, i.e., comparable to
that of the emission line regions. In addition to the above lines,
\citet{Beristain01} propose that the BC in He I lines originates
from an inner wind in some objects. In either case, it is likely
that the permitted emission line regions are located within
$\sim$0.1 AU of the star.

The above paradigm for magnetospheric mass accretion and emission
line regions has been established via observations of a number of
CTTSs, but without significant consideration of their time
variation in most (if not all) cases. However, it is well known
that emission lines and UV excess are time-variable even on a
scale of several days (see Sections 3, 4, and Appendix, and
references therein). Even so, the measurements of line-vs.-line
and line-vs.-continuum in Section 4 agree well with the present
paradigm. The upper panel of Figure \ref{corr_line_vs_line} shows
that the correlation we observed with different epochs for four
CTTSs is similar to that based on single-epoch observations of 11
T Tauri stars by \citet{Muzerolle98_opt}. In Figure
\ref{corr_line_vs_line}, a good correlation between the
\ion{Ca}{+2}, \ion{He}{+1}, and \ion{O}{+1} 8446 \AA\ and Pa 11
lines is explained if these are mainly associated with
magnetospheric accretion columns. The worse correlations with
H$\alpha$, \ion{O}{+1} 7772 \AA\ and Na D may be due to either
different excitations and/or optical thicknesses, or contributions
from the inner wind. The former explanation may apply in
particular for \ion{O}{+1} 7772 \AA, in which remarkable
redshifted absorption is observed for RY Tau and RW Aur A. The
latter explanation may apply in particular for (1) H$\alpha$, in
which the contribution from the inner wind is suggested
\citep[e.g.,][]{Takami01,Takami03, Alencar05,Kurosawa06}; and (2)
Na D lines, in which blueshifted absorption is observed for XZ Tau
and RW Aur A (Figs \ref{XZTau_profiles} and \ref{RWAur_profiles},
respectively).

In Figure \ref{corr_line_vs_cont}, the continuum luminosity at
5000 \AA\ shows good correlation with the \ion{Ca}{2} 8542 \AA\
luminosity. Relatively large variations in DG Tau, XZ Tau and RW
Aur A are explained if the UV excess (or ``veiling continuum'')
significantly contributes to the total continuum luminosity; the
constancy of the continuum in RY Tau is explained if the
contribution of veiling continuum to the total flux is negligible
compared with the stellar flux. This may have caused difficulties
in measuring the rotational period of the star using the technique
of photometric monitoring (see Appendix for details).

Stellar and veiling continuum are often separated when comparing
the stellar absorption features with template spectra of main
sequence stars \citep[e.g.,][]{Hartigan95,Muzerolle98_opt}. In
this paper we do not show the calculation of veiling analysis
since we find it difficult in particular for RW Aur A, whose
stellar absorption was fairly shallow during our observations.
Furthermore, the stellar absorption is fairly shallow compared
with the noise level at the continuum for DG Tau. This results in
large uncertainties when we measure the veiling. The veiling for
RY Tau is very small and we can only get the upper limit ($<0.02$
at $\sim6000$ \AA) of the veiling value. The only successful
measurement is for XZ Tau, and we find that the veiling for XZ Tau
is between $\sim0.5$ to 1.4 at $\sim6000$ \AA\ during the six
observation dates.

\subsection{Long-Term Variations in Line Profiles, Equivalent Widths and Luminosities}

Observations of the extended jet/outflows associated with DG Tau,
XZ Tau and RW Aur A suggest that the mass ejections from these
CTTSs are episodic. DG Tau and RW Aur A show a collimated jet, and
the spatial intervals of the knots and their proper motions
suggest the intervals of such mass ejection of 2.5--5 years for DG
Tau \citep{Pyo03,Agra11,Rodriguez12}, and 3--20 years for RW Aur A
\citep{Lopez03}. In the case of XZ Tau, the observations of the
bubble-like flows and simulations suggest the intervals of
episodic mass ejection of 5--10 years \citep{Krist08}. See
Appendix for details of these jets/outflows.

Theories suggest that energetic mass ejection is powered by mass
accretion \citep[e.g.,][]{Shu00, Konigl00}, and this has been
supported by observations, e.g., a good correlation between the
accretion luminosity vs. forbidden line luminosity associated with
the jet/wind or inferred mass ejection rate
\citep{Hartigan95,Calvet97}. This is based on the assumption of
steady mass ejection and accretion, but following this, it is
natural to assume that the episodic mass ejection suggested by the
jet/wind structures is caused by episodic mass accretion. Episodic
mass accretion is observed in some young stellar objects as FUor
or EXor bursts \citep[e.g.,][]{Herbig89}, but the link with mass
ejection has not been observationally investigated in detail. The
measurements of the time lag between episodic mass ejection and
accretion would be useful for exploring their physical link.

Episodic mass accretion associated with FUors and EXors is often
observed via photometry at optical wavelengths. However, such
optical bursts may also be observed due to sudden change in
obscuration or extinction by the dusty environments associated
with the circumstellar disk (UXors). Indeed, such events were
reported for one of our targets, RY Tau \citep[see also
Appendix]{Zajtseva85,Herbst84,Herbst94,Petrov99}. The measurements
of the change in optical emission lines, either the shape of line
profiles or equivalent widths, would be able to overcome this
problem.

However, the shape of line profiles and equivalent widths also
show variability over a few days or months --- i.e., variability
not directly related to the above issue. As shown in Sections 3
and 4, daily variability in line profiles is also observed for
H$\alpha$, \ion{Ca}{+2}, and \ion{O}{+1} lines in RY Tau, and the
redshifted side of the Na D lines in RW Tau A;  all the objects we
observed show the daily variability in the equivalent widths and
luminosities of the emission lines. Possible mechanisms for this
variability are: (1) a combination of stellar rotation and
non-axisymmetric accretion; and (2) the time variation of mass
accretion (and also mass ejection). The former should have a
period of 2--20 days, and such a periodicity in line profiles and
luminosities was observed in some CTTSs including RW Aur A
\citep{Petrov01a}, AA Tau \citep{Bouvier07_AATau}, Sz 68
\citep{Johns97} and SU Aur
\citep{Giampapa93,Johns95_SUAur,Petrov96}. In contrast,
photometric monitoring of active CTTSs shows irregular variability
in some stars even in a single rotation period, making it
difficult to determine their rotational periods (see Appendix).
Among our targets, such variability is measured in RW Aur A.
Furthermore, tidal interaction with a close binary companion could
cause periodic time variability of the mass accretion rate, and
thereby the emission line profiles, equivalent widths and
luminosities \citep[e.g.,][]{Basri97,Petrov01a}. In this context,
the variability over 2--20 days may not be always caused by a
combination of stellar rotation and non-axisymmetric accretion.

As described in Section 3.2, we find clear differences in some
line profiles between the literature and our observations. This is
consistent with the presence of long-term ($\gg$1 year), or
transient variability \citep[e.g.,][]{Oliveira00} is detectable
over the daily and monthly variabilities. Such variation may be
related to the episodic mass ejection suggested by the
high-resolution imaging. Continuous and simultaneous observations
of both optical emission lines and extended jet/wind structure are
necessary to investigate this.

\section{Summary}

The time variation of the permitted emission line profiles for
four active CTTSs in six nights of observations in 2010 is
presented in this study. The correlations of equivalent widths and
of extinction-corrected luminosities between different lines are
also investigated. We show similar correlations between our
observation in different epochs and single-epoch observations of
11 T Tauri stars by \citet{Muzerolle98_opt}. The good correlations
between certain lines suggest these lines are mainly associated
with magnetospheric mass accretion. These measurements support the
paradigm for steady magnetospheric mass accretion and the emission
line regions which are proximity to the star.

Moreover, comparing to samples in the literature, we confirm the
presence of time variability separate from the daily and monthly
variabilities during our observations. We therefore expect to
detect a variability over a time scale of $\gg$1 years due to
time-variable mass accretion in the planned observations. This
might help explore the time correlation between mass ejection seen
in the jet/wind structure and mass accretion probed by permitted
emission lines, and further understand the driving mechanism of
the jet/wind and the physical link with mass accretion. A
long-term monitoring of magnetospheric mass accretion and the
extended jets is ongoing, and the analysis of data obtained in
2011 and 2012 are in progress.

\acknowledgments

This research made use of the Simbad database operated at CDS,
Strasbourg, France, and the NASA Astrophysics Data System Abstract
Service. We appreciate helpful conversations with Chun-Fan Liu,
and useful comments from the anonymous referee. MT is supported by
the National Science Council of Taiwan (grant no.
100-2112-M-001-007-MY3).

{\it Facilities:} \facility{Canada-France-Hawaii Telescope
(ESPaDOnS)}.

\appendix

\section{Note for Individual Targets}

\subsection{DG Tau}
The DG Tau jet is one of the best studied among those associated
with CTTSs. An extended jet structure has been observed at various
wavelengths including optical
\citep{Mundt83,Mundt87,Solf93,Kepner93,Lavalley97,Eisloffel98,Lavalley00,Dougados00,Bacciotti00,Bacciotti02,Coffey07,Coffey08},
near-infrared \citep{Takami02b,Pyo03,Beck08,Agra11}, X-ray
\citep{Gudel05,Gudel08} and centimeter wavelengths
\citep{Bieging84,Rodriguez12}. In the above literature, the jet
structure was observed at an angular resolution of $\sim$0''.1 by
\citet{Kepner93,Dougados00,Bacciotti00,Bacciotti02,Pyo03,Coffey07,Coffey08,Beck08,Agra11,Rodriguez12}.
Models and theoretical work specifically for this jet have been
made by, e.g.,
\citet[][]{Raga01,Anderson03,Pesenti04,Cerqueira04,Massaglia05,Gunther09}.
Their proper motions and inferred episodic mass ejection were
extensively studied in \citet{Pyo03}, followed by
\citet{Agra11,Rodriguez12}. \citet{Agra11} derived intervals
between knots of $\sim$2.5 years between 1995 and 2005. This is
smaller by a factor of 2 than that provided by
\citet{Pyo03,Rodriguez12} ($\sim$5 years).

Neither lunar occultations nor near-infrared speckle imaging show
the evidence for a binary companion in a separation range of
0''.005--10'' \citep{Leinert91,Leinert93,Simon92,Ghez93}, but
near-infrared interferometry by \citet{Colavita03} does not
exclude a possibility of the presence of a close companion with a
separation of $\ll$0''.1.

The rotational period of the star was measured through long-term
photometric monitoring campaigns in the $UBVRI$ bands,
specifically COYOTES I and II \citep{Bouvier93,Bouvier95}. Based
on the COYOTES I data \citet{Bouvier93} measured a period of 6.3
days with 99.9\% confidence. In contrast, the COYOTES II data in
the $UBVRI$ do not clearly show any significant period
\citep{Bouvier95}. However, \citet{Bouvier95} report that the
combination of the COYOTES I and II data still show a rotational
period of 6.3 days with 95\% confidence in the $U$-band.

\subsection{RY Tau}

The extended optical jet has been observed by
\citet{St-Onge08,Agra09}. The latter provided the image of the jet
at an angular resolution of 0''.2. Near-infrared speckle imaging
observations did not show the presence of a binary companion at
0''.1--1'' \citep{Ghez93,Leinert93}. \citet{Bertout99} measured
the variability of the astrometric motion measured using Hipparcos
and interpreted it as indicative of a binary with minimum
separation of 3.3 AU. However, \citet{Agra09} point out that their
results may be due to ''UXor-like occultation events, which
enhance emission from the blueshifted jet and/or scattering cavity
relative to the occulted photosphere.'' \citet{Herbig77} suspected
that RY Tau is a spectroscopic binary with an amplitude of 25 km
s$^{-1}$, however, more recent measurement with a higher accuracy
did not confirm this \citep{Hartmann86,Petrov99}.

RY Tau has long been subject to extensive photometric monitoring.
The COYOTE I data suggest a period of $24\pm2$ days
\citep{Bouvier93}.  \citet{Bouvier93} point out that this cannot
result from a rotational model since, for a measured $v$ sin $i$
$\sim$ 50 km s$^{-1}$ \citep{Hartmann89,Petrov99}, a rotational
period of 24 days would imply a stellar radius of 24 $R_\odot$ or
larger. Other possible periods discussed to date include 5.6
(Herbst et al. 1987, but not confirmed by Herbst \& Koret 1988 and
Bouvier et al. 1993) and 66 days \citep[][but not for the rotation
period]{Herbst87}. The above three periods are close to harmonic
or sub-harmonic multiples of one another \citep{Bouvier93}. RY Tau
also shows a peculiar photometric variability with larger
variations of brightness accompanied by a near constancy of color.
Two abrupt brightening events were recorded in 1983/1984 and
1996/1997 reminiscent of UXor events, interpreted as variable
obscuration by circumstellar dust
\citep{Zajtseva85,Herbst84,Herbst94,Petrov99}.

\subsection{XZ Tau}

This object is known to host an unusual bubble-like outflow at an
angular scale of $\sim$5''
\citep{Krist97,Krist99,Krist08,Hioki09}. All of these observations
were made at an angular resolution of $\sim$0''.1. \citet{Krist08}
have shown marked changes in the brightness distribution of the
bubble through observations over a decade. Models specific for the
bubble outflow have been provided by, e.g.,
\citet[][]{Coffey04,Krist08}. Based on simulations,
\citet{Krist08} attributed the observed structure to interactions
between ambient material and a pulsed jet, whose intervals are
5-10 years. XZ Tau is a binary system with a separation of 0''.3
\citep{Haas90,Ghez93,White01,Hartigan03,Beck08}. The spectrum of
the S component  indicates that it is a typical T Tauri star,
while the N component has large spectral veiling and numerous
emission lines \citep{Hartigan03}. Recent VLA observations by
\citet{Carrasco09} have revealed that the S component is a close
binary with a separation of 0''.09.

Photometric monitoring by COYOTES II measured a periodicity of 2.6
days at a confidence level of $90\%$ \citep{Bouvier95}.
\citet{Bouvier95} stated that it may imply too large equatorial
velocity of the star (46.3 km s$^{-1}$) compared with those of
active T Tauri stars \citep[usually $<$20 kms$^{-1}$,
e.g.,][]{Hartmann89}, but it is comparable to that measured in RY
Tau \citep[$52 \pm 2$ km s$^{-1}$,][]{Petrov99}. Ground-based
photometry from 1962 to 1981 \citep{Herbst94} has shown variations
of $\Delta m \sim2$ mag in the $V$ band. All of these photometric
measurements were made for the binary as a whole.

Variability of each binary component has been monitored using the
{\it Hubble Space Telescope} over 10 years in $R$-band
\citep[1995--2004,][]{Krist08}. \citet{Coffey04,Krist08} reported
a dramatic brightening ($\Delta m \sim$3 mag) of XZ Tau N in
1998--2002, suggesting that it is an EXor-type variable
\citep{Herbig89}. In 2001, its $R$ magnitude was larger than XZ
Tau S by 0.5 mag. \citet{Krist08} shows that XZ Tau S was far less
variable ($\Delta m \sim$ 0.5 mag) during the period of
monitoring, and it was brighter than XZ Tau N by 1--3 mag except
in 2001. In addition to the optical wavelengths, variability in
$H$-band (1.65 $\micron$) was also observed over a 3--15 year
period \citep{Hioki09}.

\subsection{RW Aur A}

Similar to DG Tau, there are a number of observations for the
extended jet associated with RW Aur A at optical
\citep{Hirth94,Bacciotti96,Mundt98,Dougados00,Woitas02,Woitas05,Lopez03,Coffey04,Coffey12,Melnikov09,Liu12}
and near-infrared wavelengths
\citep{Davis03,Pyo06,Beck08,Hartigan09}. Among them,
\citet{Dougados00,Woitas02,Pyo06,Beck08,Melnikov09,Hartigan09}
show the jet structure at an angular resolution of $\sim$0''.1. As
many of these papers show, RW Aur A is associated with a jet on
both blueshifted and redshifted sides, with the latter brighter
than the former. This contrasts with many other CTTSs, in which
the redshifted emission is absent close to the star due to
obscuration by a circumstellar disk \citep[see][for a
review]{Eisloffel00}. Many of above publications also show that
the kinematics of the RW Aur A jet are asymmetric, measured at
$\sim-200$ and $\sim100$ km s$^{-1}$, respectively. \citet{Liu12}
interpret such an asymmetry is a result of asymmetric launching of
the wind by the underlying magnetic field that is asymmetric on
the opposite side of the disk, in a way that the linear momentum
is conserved. Moreover, \citet{Lopez03} measured a proper motion
of the RW Aur A jet of 0''.17--0''.27 yr$^{-1}$ and $\sim$0''.26
for the redshifted and blueshifed sides of the jet, respectively.
These authors propose that the distribution of the knots close to
the star is explained by time variable mass ejection with at least
two modes: one irregular and asymmetric on timescale of $\le$3--10
years, and another more regular with an $\sim20$-year period.

RW Aur A has a companion with a separation of 1''.4
\citep{Ghez93,Leinert93,Dougados00,White01}. The jet described
above is associated with RW Tau A \citep{Dougados00}. RW Aur A is
also  a suspected spectroscopic binary \citep{Petrov01b}.

RW Aur A shows a complicated variability in optical continuum,
emission and absorption lines. \citet{Herbst94} classified this
star as ''irregular variability'' (type II) through their
monitoring observations of the optical continuum over 20 years.
\citet{Gahm99,Petrov01a,Petrov01b} observed variations in radial
velocity and emission/absorption line properties over 2.77- and
5.5-day periods. These authors attributed this to a close low-mass
companion or to a rotationally modulated accretion ``hot spot''.
In contrast, \citet{Alencar05} observed, in bright emission lines,
2.77- and 3.9-day periods in their data in 1992, and 4.2- and
5.5-day periods in 1999. Unlike RY Tau, no abrupt brightening is
identified over 20 years of photometry, combined by
\citet{Herbst94}.

\bibdata{astro.bib}

\bibliography{astro}

\clearpage

\clearpage

\clearpage

\clearpage

\clearpage

\begin{table}
\begin{center}

\caption{$VRI$ Magnitudes of Standard Stars \label{std_BVRI}}
\begin{tabular}{lccc}
\tableline\tableline
Star & \multicolumn{3}{c}{Magnitude} \\
& $V$ & $R$ & $I$ \\ \tableline \tableline
HD 283642& 10.60 & 10.06 & 9.52\\
HD 42784 & 6.53 & 6.60 & 6.64\\

\tableline
\end{tabular}
\end{center}
\end{table}

\clearpage

\begin{deluxetable}{lllclccc}
\rotate \tabletypesize{\tiny} \tablecaption{Previous observations
of optical permitted line profiles \label{literature}}
\tablewidth{0pt} \tablehead{ \colhead{Object} & \colhead{Year of}
& \colhead{Line(s)} & \colhead{Spectral resolution} &
\colhead{Reference} & \colhead{Study of time} &
\colhead{Equivalent width} & \colhead{Note}
\\
\colhead{}  & \colhead{observations} & \colhead{} & \colhead{$R /
10^4$} & \colhead{} & \colhead{variability} & \colhead{in
H$\alpha$ (\AA)} & \colhead{} }

\startdata
DG Tau
        & 1981
            & H$\alpha$, Na D
            & 2.5
            & \citet{Mundt84}
            &
            & 110
                    \\

        & 1987
            & \ion{Ca}{+2} 8498/8542 \AA, \ion{O}{+1} 7772/8446 \AA\
            & 1.4
            & \citet{Hamann92a}
            &
            &
            &   \\

        & 1988--1992
            & H$\alpha$, \ion{He}{+1} 5876/6678 \AA\
            & 2.5
            & \citet{Beristain01}
            & Y
            &
            & a \\

        & 1996
            & H$\alpha$, Pa 11, \ion{Ca}{+2} 8542 \AA, \ion{He}{+1} 5876 \AA, \ion{O}{+1} 7772/8446 \AA, Na D
            & 3.5
            & \citet{Muzerolle98_opt}
            &
            & 46.2
            \vspace{0.3cm}\\

RY Tau
        & 1980
            & H$\alpha$
            & 0.2--0.3
            & \citet{Mundt82}\\

        & 1984
            & H$\alpha$
            & 0.2--0.3
            & \citet{Zajtseva85}
            & Y
            & 8.5--20
            & b\\

        & 1984
            & H$\alpha$
            & 2.5
            & \citet{Edwards87}\\

        & 1987
            & \ion{Ca}{+2} 8498/8542 \AA, \ion{O}{+1} 7772/8446 \AA\
            & 1.4
            & \citet{Hamann92a}     \\

        & 1987--1988
            & H$\alpha$, \ion{He}{+1} 5876 \AA, Na D
            & 1.6
            & \citet{Petrov90}
            & Y
            &
            & c\\

        & 1989--1996
            & H$\alpha$, \ion{Ca}{+2} triplet, Na D
            & 2.5/4.0
            & \citet{Petrov99}
            & Y
            & 5.9--24.9
            & d\\

        & 1992
            & H$\alpha$
            & 4.8
            & \citet{Johns95}
            &Y
            &
            & e\\

        & 1998--1999
            & H$\alpha$, \ion{He}{+1} 5876 \AA, Na D
            & 0.55
            & \citet{Mendigutia11a}
            & Y
            & 15.3$\pm$0.2
            & f \vspace{0.3cm}\\

XZ Tau
        & 1983, 1987
            & H$\alpha$, \ion{Ca}{+2} 8498/8542 \AA, \ion{O}{+1} 7772/8446 \AA\
            & 0.35/1.4
            & \citet{Hamann92a}
            &
            &
            & g \vspace{0.3cm}\\

RW Aur A
        & 1980
            & H$\alpha$, Na D
            & 2.5
            & \citet{Mundt84}
            &
            & 85
            & h \\

        & 1987
            & \ion{Ca}{+2} 8498/8542 \AA, \ion{O}{+1} 7772/8446 \AA\
            & 1.4
            & \citet{Hamann92a}     \\

        & 1988--1992
            &H$\alpha$, \ion{He}{+1} 5876/6678 \AA\
            & 2.5
            & \citet{Beristain01}
            &
            &
            & \\

        & 1989--1990
            & H$\alpha$
            & $\sim 1$
            & \citet{Fernandez95}
            & Y
            &71.3, 122.7, 94.7, 52.8
            & i\\

        & 1989-1993, 1999
            & H$\alpha$, \ion{Ca}{+2} 8498 \AA, \ion{He}{+1} 5876 \AA, Na D
            & 4.8
            & \citet{Alencar05}
            & Y
            &
            & j\\

        & 1995--1999
            & H$\alpha$, Pa 11, \ion{Ca}{+2} 8498 \AA, \ion{He}{+1} 5876 \AA, \ion{O}{+1} 7772 \AA. Na D
            & 2.6
            & \citet{Petrov01a}
            & Y
            &
            & k\\

        & 1996
            & H$\alpha$, Pa 11, \ion{Ca}{+2} 8542 \AA, \ion{He}{+1} 5876 \AA, \ion{O}{+1} 7772/8446 \AA, Na D
            & 3.5
            & \citet{Muzerolle98_opt}
            &
            & 50.3
            \\

\enddata

\tablenotetext{a}{3 visits for H$\alpha$ and \ion{He}{+1} 5876
\AA. Line-to-continuum ratio is shown for only one of the
\ion{He}{+1} 5876 \AA\ profiles.} \tablenotetext{b}{8 visits, with
intervals ranging from 1 day to 2 months}
\tablenotetext{c}{According to the authors 21 visits were made,
but a limited number of profiles are shown. These are 3 visits for
H$\alpha$, with an interval of 1 and 2 months; 7 visits for
\ion{He}{+1} and Na D lines, with intervals ranging from 1 day to
2.5 months.} \tablenotetext{d}{19 visits, with intervals ranging
from 2 hr to 6 years. Two different spectrographs were used (SOFIN
at the 2.56 m Nordic Optical Telescope, the coud\'e spectrograph
at the 2.6-m Shajn refrector). The paper referred to the
\ion{Ca}{+2} lines as ``\ion{Ca}{+2} triplet'', and does not
describe any more details. } \tablenotetext{e}{50 visits in 3
months. Individual profiles are not shown (only the average and
variance profiles are shown).} \tablenotetext{f}{6 visits (October
25, 26, 27, 28, 1998 and January 29, 30, 1999). Individual
profiles are not shown (only the average and variance profiles are
shown).} \tablenotetext{g}{H$\alpha$ was observed in 1983 at $R =
3.5 \times 10^3$, while the others were observed in 1987 at $R =
1.4 \times 10^4$. } \tablenotetext{h}{The same profiles are also
shown in \citet{Mundt82}. } \tablenotetext{i}{Equivalent widths
are shown for 4 visits, but the line profile is shown for only one
of them.} \tablenotetext{j}{77 visits in total, with intervals
ranging from 1 hour to 1 year.} \tablenotetext{k}{38 visits, with
intervals ranging from 1 hour to 1 year. The same data were also
used in \citet{Gahm99,Petrov01a,Petrov07}.}

\end{deluxetable}

\begin{deluxetable}{lrrrrrrrrrrrrrr}
\tabletypesize{\tiny} \tablecaption{Equivalent
Widths\tablenotemark{1} \label{EWs}} \tablewidth{0pt} \tablehead{
\colhead{Object} & \colhead{\ion{He}{+1}} & \colhead{Na D} &
\colhead{H$\alpha$} & \colhead{\ion{He}{+1}} &
\colhead{\ion{He}{+1}} & \colhead{\ion{O}{+1}}
&\colhead{\ion{O}{+1}}& \colhead{\ion{Ca}{+2}} &
\colhead{\ion{Ca}{+2}} & \colhead{\ion{Ca}{+2}} &\colhead{Pa 11}
&\colhead{[\ion{O}{+1}]}& \colhead{[\ion{S}{+2}]} &
\colhead{[\ion{Fe}{+2}]}
\\
\colhead{UT Date} & \colhead{$\lambda$5876} & \colhead{} &
\colhead{$\lambda$6563} & \colhead{$\lambda$6678} &
\colhead{$\lambda$7065} & \colhead{$\lambda$7772} &
\colhead{$\lambda$8446} &\colhead{$\lambda$8498}&
\colhead{$\lambda$8542} & \colhead{$\lambda$8662} &
\colhead{$\lambda$8862} & \colhead{$\lambda$6300} &
\colhead{$\lambda$6731} & \colhead{$\lambda$7155}
 }

\startdata

DG Tau &  &  &  &  &  & &  &  &  & & &  & & \\
2010 Oct 21 & 2.9 & 0.8 & 87.7 & 1.2 & 0.5 & 1.8 & 3.0 & 37.3 & 38.7 & 33.1 & 3.2 & 12.4 & 1.7 & 0.5\\
2010 Nov 17 & 3.2 & 1.6 & 88.4 & 1.2 & 0.7 & 1.7 & 3.0 & 47.2 & 52.3 & 46.1 & 4.1 & 10.2 & 1.5 & 0.4 \\
2010 Nov 21 & 3.1 & 0.7 & 74.2 & 1.3 & 0.8 & 1.1 & 2.3 & 45.0 & 51.8 & 44.5 & 4.7 & 8.4 & 1.2 & 0.4\\
2010 Nov 25 & 3.2 & 0.4 & 87.1 & 1.2 & 0.9 & 1.2 & 2.9 & 47.6 & 50.9 & 43.7 & 3.9 & 9.4 & 1.3 & 0.4\\
2010 Nov 27 & 3.3 & 0.9 & 106.9 & 1.1 & 0.7 & 1.8 & 3.8 & 54.4 & 55.9 & 48.0 & 4.8 & 10.6 & 1.4 & 0.4\\
2010 Dec 17 & 3.0 & -0.7 & 102.2 & 1.4 & 0.5 & 2.0 & 2.9 & 35.0 & 34.8 & 30.3 & 3.1 & 14.8 & 2.1 & 0.8\\
\\
RY Tau &  &  &  &  &  & &  &  &  & & &  & & \\
2010 Oct 21 & 0.2 & -1.4 & 10.3 & -0.03 & ...\tablenotemark{2} & 0.2 & 1.0 & 2.9 & 1.1 & 1.3 & -0.4 & 0.0 & ...\tablenotemark{2} & ...\tablenotemark{2} \\
2010 Nov 17 & 0.2 & -0.9 & 14.9 & -0.01 & ... & 0.3 & 1.4 & 5.6 & 3.7 & 3.6 &-0.3 & 0.1 & ... & ... \\
2010 Nov 21 & 0.4 & -1.0 & 15.5 & -0.02 & ... & 0.2 & 1.0 & 3.6 & 2.1 & 2.2 &-0.3 & 0.1 & ... & ... \\
2010 Nov 25 & 0.0 & -1.4 & 14.3 & -0.07 & ... & -0.4 & 0.9 & 3.1 & 1.6 & 1.7 &-0.7 & 0.0 & ... & ... \\
2010 Nov 27 & -0.2 & -2.1 & 11.8 & -0.14 & ... & -1.1 & 0.4 & 3.1 & 1.4 & 1.5 &-0.8 & 0.0 & ... & ... \\
\\
XZ Tau &  &  &  &  &  &  &  &  &  & & &  & & \\
2010 Oct 21 & 3.0 & 3.6 & 166.8 & ...\tablenotemark{3} & 0.9 & ...\tablenotemark{4} & 3.2 & 34.1 & 36.4 & 27.8 & ...\tablenotemark{3} & 6.3 & 1.6 & 0.7\\
2010 Nov 17 & 4.0 & 4.2 & 160.2 & ... & 0.9 & ... & 2.5 & 31.9 & 32.6 & 25.7 & ... & 7.9 & 2.1 & 0.8\\
2010 Nov 21 & 3.0 & 5.3 & 167.4 & ... & 0.9 & ... & 2.5 & 32.8 & 33.6 & 26.1 & ... & 10.7 & 3.0 & 0.9\\
2010 Nov 25 & 2.5 & 2.7 & 173.5 & ... & 0.7 & ... & 1.8 & 23.2 & 22.8 & 17.7 & ... & 13.8 & 4.0 & 1.2\\
2010 Nov 27 & 2.4 & 4.6 & 170.0 & ... & 0.7 & ... & 1.9 & 25.4 & 26.6 & 20.5 & ... & 13.2 & 4.0 & 1.1\\
2010 Dec 17 & 1.7 & -0.1 & 158.2 & ... & 0.6 & ... & 0.4 & 9.5 & 9.5 & 7.0 & ... & 16.1 & 5.2 & 1.5\\
\\
RW Aur A&  &  &  &  &  &  & &  &  & & &  & & \\
2010 Nov 17 & 1.8 & -2.2 & 80.4 & 1.8 & 1.0 & -1.1 & 6.4 & 47.6 & 44.7 & 35.8 & 3.9 & 2.5 & 0.7 & 1.1\\
2010 Nov 21 & 1.6 & -2.8 & 65.6 & 1.3 & 0.7 & -1.2 & 5.1 & 46.8 & 46.0 & 36.6 & 3.3 & 2.8 & 0.7 & 0.7\\
2010 Nov 25 & 2.0 & -0.9 & 79.7 & 1.6 & 1.0 & -1.4 & 8.1 & 70.4 & 71.9 & 59.1 & 3.2 & 4.1 & 1.0 & 1.8\\
2010 Nov 27 & 1.7 & -0.3 & 72.9 & 1.4 & 0.8 & -1.6 & 6.9 & 70.9 & 71.7 & 58.2 & 3.3 & 6.2 & 1.6 & 2.5\\

\enddata

\tablenotetext{1}{In Angstroms. Positive values denote emission,
and negative values denote absorption.} \tablenotetext{2}{These is
no clearly detected emission or absorption feature.}
\tablenotetext{3}{Although we clearly detect the emission feature
(Fig. \ref{XZTau_profiles}), accurate measurements of the
equivalent widths were not possible due to ambiguity of the
adjacent continuum.} \tablenotetext{4}{These are excluded from the
measurement of equivalent widths as the line profiles are covered
with features whose origin is presumably different.}

\end{deluxetable}

\clearpage

\begin{deluxetable}{llrrrrrrrrrrrrrrrr}
\rotate\tabletypesize{\tiny} \tablecaption{Extinction-Corrected
Line Luminosities ($L_\odot$)\tablenotemark{1}
\label{line_luminosities}} \tablewidth{0pt} \tablehead{
\colhead{Object} & \colhead{$A_V$} & \colhead{\ion{He}{+1}} &
\colhead{Na D} & \colhead{H$\alpha$} & \colhead{\ion{He}{+1}} &
\colhead{\ion{He}{+1}} & \colhead{\ion{O}{+1}}
&\colhead{\ion{O}{+1}}& \colhead{\ion{Ca}{+2}} &
\colhead{\ion{Ca}{+2}} & \colhead{\ion{Ca}{+2}} &\colhead{Pa 11} &
\colhead{[\ion{O}{+1}]} &\colhead{[\ion{S}{+2}]}&
\colhead{[\ion{Fe}{+2}]} &\colhead{continuum} &\colhead{continuum}
\\
\colhead{UT Date} & \colhead{} & \colhead{$\lambda$5876} &
\colhead{} & \colhead{$\lambda$6563} & \colhead{$\lambda$6678} &
\colhead{$\lambda$7065} & \colhead{$\lambda$7772} &
\colhead{$\lambda$8446} &\colhead{$\lambda$8498}&
\colhead{$\lambda$8542} & \colhead{$\lambda$8662} &
\colhead{$\lambda$8862} & \colhead{$\lambda$6300} &
\colhead{$\lambda$6731} & \colhead{$\lambda$7155}
&\colhead{$\lambda$5000} & \colhead{$\lambda$8000}
 }

\startdata
DG Tau & 2.2\tablenotemark{2} &  & &  & &  &  &  &  & &  &  & & & & & \\
2010 Oct 21 && 3.4E-04 & 9.7E-05 & 1.0E-02 & 1.4E-04 & 5.4E-05 & 1.9E-04 & 3.4E-04 & 3.6E-03 & 3.9E-03 & 3.2E-03 & 3.1E-04 & 1.4E-03 & 1.9E-04 & 5.8E-05 & 1.3E-04 & 1.1E-04\\
2010 Nov 17 && 5.0E-04 & 2.4E-04 & 1.4E-02 & 1.9E-04 & 1.2E-04 & 2.7E-04 & 4.6E-04 & 6.3E-03 & 6.9E-03 & 6.1E-03 & 5.0E-04 & 1.7E-03 & 2.3E-04 & 7.6E-05 & 1.8E-04 & 1.4E-04\\
2010 Nov 21 && 6.4E-04 & 1.4E-04 & 1.5E-02 & 2.6E-04 & 1.7E-04 & 2.2E-04 & 4.6E-04 & 7.8E-03 & 8.6E-03 & 7.5E-03 & 7.0E-04 & 1.7E-03 & 2.5E-04 & 8.5E-05 & 2.5E-04 & 1.9E-04\\
2010 Nov 25 && 5.8E-04 & 7.7E-05 & 1.5E-02 & 2.2E-04 & 1.4E-04 & 2.0E-04 & 4.6E-04 & 6.9E-03 & 7.2E-03 & 6.1E-03 & 5.0E-04 & 1.7E-03 & 2.3E-04 & 6.7E-05 & 2.1E-04 & 1.6E-04\\
2010 Nov 27 && 5.2E-04 & 1.4E-04 & 1.6E-02 & 1.8E-04 & 1.0E-04 & 2.8E-04 & 5.5E-04 & 7.2E-03 & 7.2E-03 & 6.1E-03 & 5.8E-04 & 1.7E-03 & 2.3E-04 & 7.6E-05 & 1.7E-04 & 1.5E-04\\
2010 Dec 17 && 3.8E-04 & ... & 1.4E-02 & 1.9E-04 & 6.7E-05 & 2.7E-04 & 4.3E-04 & 4.5E-03 & 4.5E-03 & 3.8E-03 & 3.9E-04 & 2.0E-03 & 2.8E-04 & 1.1E-04 & 1.3E-04 & 1.4E-04\\
\\
RY Tau & 2.2\tablenotemark{3} &  &  & &  &  & &  &  & &  &  & & & & & \\
2010 Oct 21 && 4.2E-04 & ... & 1.6E-02 & ... & ... & 2.0E-04 & 1.0E-03 & 2.9E-03 & 1.1E-03 & 1.2E-03 & ... & 5.3E-05 & ... & ... & 2.5E-03 & 1.1E-03\\
2010 Nov 17 && 3.2E-04 & ... & 2.1E-02 & ... & ... & 3.5E-04 & 1.3E-03 & 4.8E-03 & 3.0E-03 & 2.9E-03 & ... & 8.4E-05 & ... & ... & 2.3E-03 & 9.6E-04\\
2010 Nov 21 && 7.8E-04 & ... & 2.4E-02 & ... & ... & 2.7E-04 & 1.0E-03 & 3.6E-03 & 2.1E-03 & 2.1E-03 & ... & 8.9E-05 & ... & ... & 2.5E-03 & 1.1E-03\\
2010 Nov 25 && 1.3E-04 & ... & 2.0E-02 & ... & ... & ... & 8.5E-04 & 2.6E-03 & 1.3E-03 & 1.4E-03 & ... & 4.1E-05 & ... & ... & 2.3E-03 & 9.9E-04\\
2010 Nov 27 && ... & ... & 1.6E-02 & ... & ... & ... & 3.4E-04 & 2.7E-03 & 1.2E-03 & 1.3E-03 & ...& ... & ... & ... & 2.3E-03 & 9.9E-04\\
\\
XZ Tau & 1.40\tablenotemark{4} &  & & &  &  &  &  &  & &  & & & & & &  \\
2010 Oct 21 && 1.0E-04 & 1.1E-04 & 7.1E-03 & ... & 3.6E-05 & ... & 2.2E-04 & 2.2E-03 & 2.4E-03 & 2.0E-03 & ... & 3.0E-04 & 8.5E-05 & 2.6E-05 & 3.3E-05 & 6.8E-05\\
2010 Nov 17 && 1.1E-04 & 1.0E-04 & 6.2E-03 & ... & 3.6E-05 & ... & 1.7E-04 & 2.0E-03 & 2.0E-03 & 1.8E-03 & ... & 3.4E-04 & 1.1E-04 & 2.6E-05 & 2.7E-05 & 6.6E-05\\
2010 Nov 21 && 5.6E-05 & 8.8E-05 & 4.8E-03 & ... & 2.6E-05 & ... & 1.5E-04 & 1.8E-03 & 1.8E-03 & 1.6E-03 & ... & 3.4E-04 & 1.1E-04 & 2.5E-05 & 1.7E-05 & 5.5E-05\\
2010 Nov 25 && 3.6E-05 & 3.3E-05 & 3.7E-03 & ... & 1.7E-05 & ... & 8.5E-05 & 1.0E-03 & 9.8E-04 & 8.7E-04 & ... & 3.1E-04 & 1.1E-04 & 2.4E-05 & 1.2E-05 & 4.4E-05\\
2010 Nov 27 && 3.6E-05 & 6.2E-05 & 4.0E-03 & ... & 1.7E-05 & ... & 9.8E-05 & 1.2E-03 & 1.2E-03 & 1.1E-03 & ... & 3.3E-04 & 1.2E-04 & 2.6E-05 & 1.4E-05 & 4.6E-05\\
2010 Dec 17 && 2.5E-05 & ... & 3.4E-03 & ... & 1.3E-05 & ... & 1.7E-05 & 4.2E-04 & 4.2E-04 & 3.5E-04 & ... & 3.6E-04 & 1.4E-04 & 2.6E-05 & 1.2E-05 & 4.6E-05\\
\\
RW Aur A& 0.39\tablenotemark{5} &  & &  &  &  &  &  &  & &  &  & & & & & \\
2010 Nov 17 && 3.8E-05 & ... & 1.6E-03 & 3.5E-05 & 1.8E-05 & ... & 1.1E-04 & 8.1E-04 & 7.4E-04 & 6.0E-04 & 5.4E-05 & 6.1E-05 & 1.6E-05 & 2.2E-05 & 2.7E-05 & 1.9E-05\\
2010 Nov 21 && 7.4E-05 & ... & 3.1E-03 & 6.0E-05 & 3.3E-05 & ... & 2.1E-04 & 2.1E-03 & 1.9E-03 & 1.6E-03 & 1.2E-04 & 1.5E-04 & 3.4E-05 & 3.0E-05 & 6.0E-05 & 4.4E-05\\
2010 Nov 25 && 5.9E-05 & ... & 2.3E-03 & 4.4E-05 & 2.6E-05 & ... & 1.8E-04 & 1.7E-03 & 1.7E-03 & 1.3E-03 & 6.2E-05 & 1.4E-04 & 2.8E-05 & 5.1E-05 & 3.9E-05 & 2.5E-05\\
2010 Nov 27 && 4.0E-05 & ... & 1.9E-03 & 3.1E-05 & 1.7E-05 & ... & 1.3E-04 & 1.3E-03 & 1.3E-03 & 1.1E-03 & 5.2E-05 & 1.6E-04 & 3.9E-05 & 5.7E-05 & 3.3E-05 & 2.1E-05\\

\enddata

\tablenotetext{1}{Those with negative equivalent widths are not
included.}\tablenotetext{2}{From \citet{Calvet98}.
}\tablenotetext{3}{From \citet{Calvet04}}\tablenotetext{4}{From
\citet{Hartigan03}}\tablenotetext{5}{From \citet{White01}}
\end{deluxetable}

\end{document}